\documentclass[aps,prl,floatfix,twocolumn,showpacs]{revtex4-2}
\usepackage{amsmath,amsthm,amssymb}
\usepackage{graphicx}
\usepackage{float}
\usepackage{siunitx}
\usepackage[utf8]{inputenc}
\usepackage{hyperref}
\begin{document}

\title{Observation of quantum oscillations in the extreme weak anharmonic limit}

\author{A. Th\'ery${}^{1}$,  B. Neukelmance${}^{1}$, B. Hue${}^{1}$, W. Legrand${}^{1}$, L. Jarjat${}^{1}$, J. Craquelin${}^{1}$, M. Villiers${}^{1}$, A. Cottet${}^{1,2}$, M.R. Delbecq${}^{1,2,3}$ and T. Kontos${}^{1,2}$\footnote{To whom correspondence should be addressed: takis.kontos@ens.fr}}
\affiliation{${}^{1}$ Laboratoire de Physique de l'\'{E}cole Normale Sup\'{e}rieure, ENS, Universit\'{e} PSL, CNRS, Sorbonne Universit\'{e}, Universit\'{e} Paris-Diderot, Sorbonne Paris Cit\'{e}, Paris, France.}
\affiliation{$^{2}$Laboratoire de Physique et d'Etude des Mat\'eriaux, ESPCI Paris, PSL University, CNRS, Sorbonne Universit\'e, Paris, France\\}
\affiliation{$^{3}$Institut universitaire de France (IUF)}

\pacs{73.23.-b,73.63.Fg}

\begin{abstract}
We investigate a granular aluminium quantum circuit with an anharmonicity of the order of its decoherence rate in a 3-dimensional microwave cavity. We perform single qubit-like manipulations such as Rabi oscillations and Ramsey fringes. Our findings, supported by quantitative numerical modeling, show that a very weakly anharmonic oscillator can also display quantum oscillations outside the qubit regime. These oscillations are hard to disambiguate from qubit oscillations in time domain measurements for a single driving frequency. This sheds new light on recent findings for new material superconducting quantum bits. Our platform shows in addition large magnetic field resilience which could find applications for quantum enhanced dark matter search.
\end{abstract}

\date{\today}
\maketitle

Superconducting circuits have become a major platform for quantum information processing and quantum amplification. Whereas the conventional Josephson junction has been extensively used so far, many efforts aim to replace it by more elaborated materials which range from semiconducting nanowires \cite{Casparis:21} or nanotubes \cite{Mergenthaler:21} to 2D materials \cite{Kroll:18,Wang:19,Wang:22}. Such setups would offer an enhanced electrical tunability (via e.g. gate electrodes) and magnetic field resilience. Both these aspects could be crucial regarding new applications for quantum sensing or for scaling up quantum information processing platforms \cite{Kroll:18,Wang:19,Wang:22}. In particular, the potential for quantum sensing applications range from quantum amplification of cosmological signals to read-out of topological systems.

These new platforms are appealing in terms of technology, but they often display a reduced anharmonicity which could endanger their qubit character. This motivates the detailed study of the dynamics of such systems at the frontier between the harmonic and the anharmonic regime, i.e.  when the anharmonicity is comparable to the damping rate of these oscillators. In this context, the use of a simple superconducting material offering both low anharmonicity and weak losses without the inherent complexity of low dimensional materials would be appealing. Granular aluminium (grAl) is such a material. It can be used both in the extreme weak anharmonic regime and in the qubit regime \cite{Maleeva:18,Pop:19,Winkel:20,Schon:20,Rieger:23}. It is therefore ideal to study the above mentionned crossover.

\begin{figure}[!ht]
\centering\includegraphics[height=0.65\linewidth,angle=0]{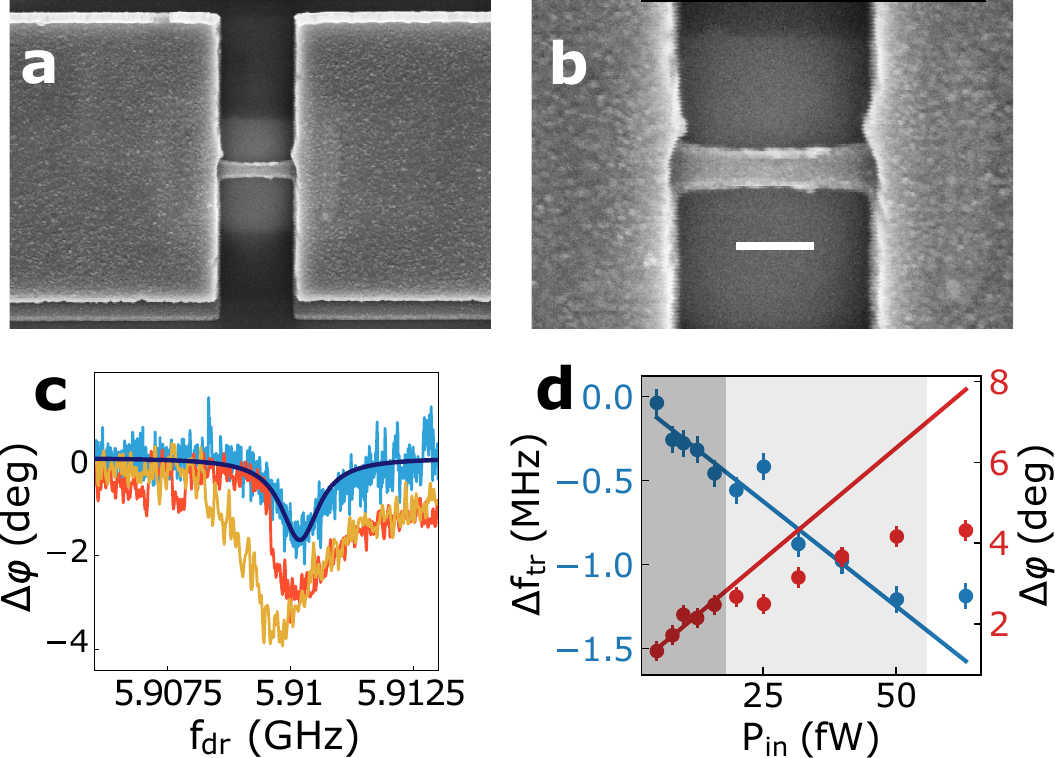}
\caption{\textbf{a.} Large scale view of the grAl transmon circuit \textbf{b.} Micrograph of the grAl Josephson inductor. The bar is \qty{250}{\nano\m}  \textbf{c.} Microwave spectroscopy of the grAl circuit for $-113$ dBm (blue), $-101$ dBm (red) and $-97$ dBm (orange). It shows the characteristic features of an anharmonic oscillator. \textbf{d.} From the power dependence of the phase contrast and the grAl circuit frequency, we extract an anharmonicity of $\approx \qty{250}{\kilo\hertz}$. The shaded dark grey region is the linear regime from which we extract these values. In this region, we can compute the excitation number of the device. We find that the device remains in the linear regime up to an average occupation number around 2.
}
\label{grAl transmon}%
\end{figure}

In this work, we use a grAl based quantum circuit in the extreme weak anharmonic regime. We show that we can perform qubit-like manipulations such as Rabi oscillations and Ramsey interferometry. However, the dynamics of the circuit involve many oscillator levels. This behavior is understood quantitatively by a simple quantum model of an anharmonic oscillator. By the analysis of the Rabi chevron patterns and the Ramsey fringes, we can assess about the "quantumness" of the observed oscillations. We show in particular that single time domain traces in these systems can easily mimic qubit dynamics both for Rabi and Ramsey oscillations. Doing so, we benchmark the time domain oscillations in the weak anharmonic regime. In addition, we are able to drive our circuit up to \qty{0.7}{\tesla}, making our simple grAl circuit appealing for magnetic field resilient quantum sensing applications.

The Josephson junction is mounted in a copper 3-dimensional (3D) microwave cavity with a fundamental frequency of about $\omega_\mathrm{cav} = 2 \pi \times \qty{6.0}{\giga\hertz}$ and a quality factor $Q=2700$ at low temperature. The whole setup is mounted in a dilution refrigerator with a standard microwave setup \cite{Bruhatthesis:16} and all the measurements are carried out at $T \approx \qty{20}{\milli\kelvin}$. Such a geometry is that of a "3D-transmon" \cite{Paik:11,Winkel:20}. The details of the nanofabrication are given in the Supplementary. From the DC switching current measurements, we estimate the supercurrent as shown in \cite{Friederich:19} to $I_c \approx \qty{600}{\nano\ampere}$(see Supplementary). The inductance of the circuit is estimated to $L_J \approx \qty{1.63}{\nano\henry}$ from the Mattis-Bardeen formula \cite{Winkel:20}. From microwave simulations, we estimate  for the layout of figure \ref{grAl transmon}, which yields a charging energy of $E_C \approx \qty{47}{MHz}$, which yields a resonance frequency of \qty{6.14}{\giga\hertz}, close to our measurements. In addition, from the DC measurements stated above, we can expect a transmon anharmonicity of $K_0 \approx \mathcal{C} \pi a \omega_{1}^{2}/j_c V_{grAl}\approx \qty{250}{kHz}$, with $\mathcal{C}=3/16$ and $a=\qty{5}{\nano\meter}$\cite{Winkel:20,Maleeva:18}, $a$ being the Al grain size. The parameters of our circuit can be determined experimentally by the microwave two-tone spectroscopy which is shown in figure \ref{grAl transmon}c, detuned from the cavity mode. We observe a clear dip at \qty{5.9102}{\giga\hertz} which displays the characteristic asymmetric lineshape of an anharmonic oscillator as the power of the microwave input signal is increased. The linewidth at lowest power is $\Gamma_{exp}\approx 2\pi \times \qty{0.935}{\mega\hertz} \pm \qty{0.1}{\mega\hertz}$, as fitted by the black line. From the shift towards low frequencies of the edge of the transmon resonance and the phase contrast as the power is increased shown in figure \ref{grAl transmon}d, we calibrate the microwave power and estimate a Kerr anharmonicity of about $K_{exp} \approx 2\pi \times \qty{250}{\kilo\hertz}$ and a coupling between the cavity and the transmon of about $g\approx\qty{16}{\mega\Hz}$. Details on this calculation can be found in the supplementary material \cite{suppmat}. This is in good agreement with the expected anharmonicity given the inherent uncertainty on the grain size of the GrAl. The linewidth observed at lowest power is directly related to the decoherence rate of the circuit. This places our circuit in the regime of extreme weak anharmonicity where $K_{exp} \lesssim \Gamma_{exp}$. Does this regime enable quantum oscillations such as Rabi-like oscillations and Ramsey fringes which are usual signatures of driven quantum systems? This question is addressed in the subsequent part of the paper.

\begin{figure}[!ht]
\centering\includegraphics[height=0.8\linewidth,angle=0]{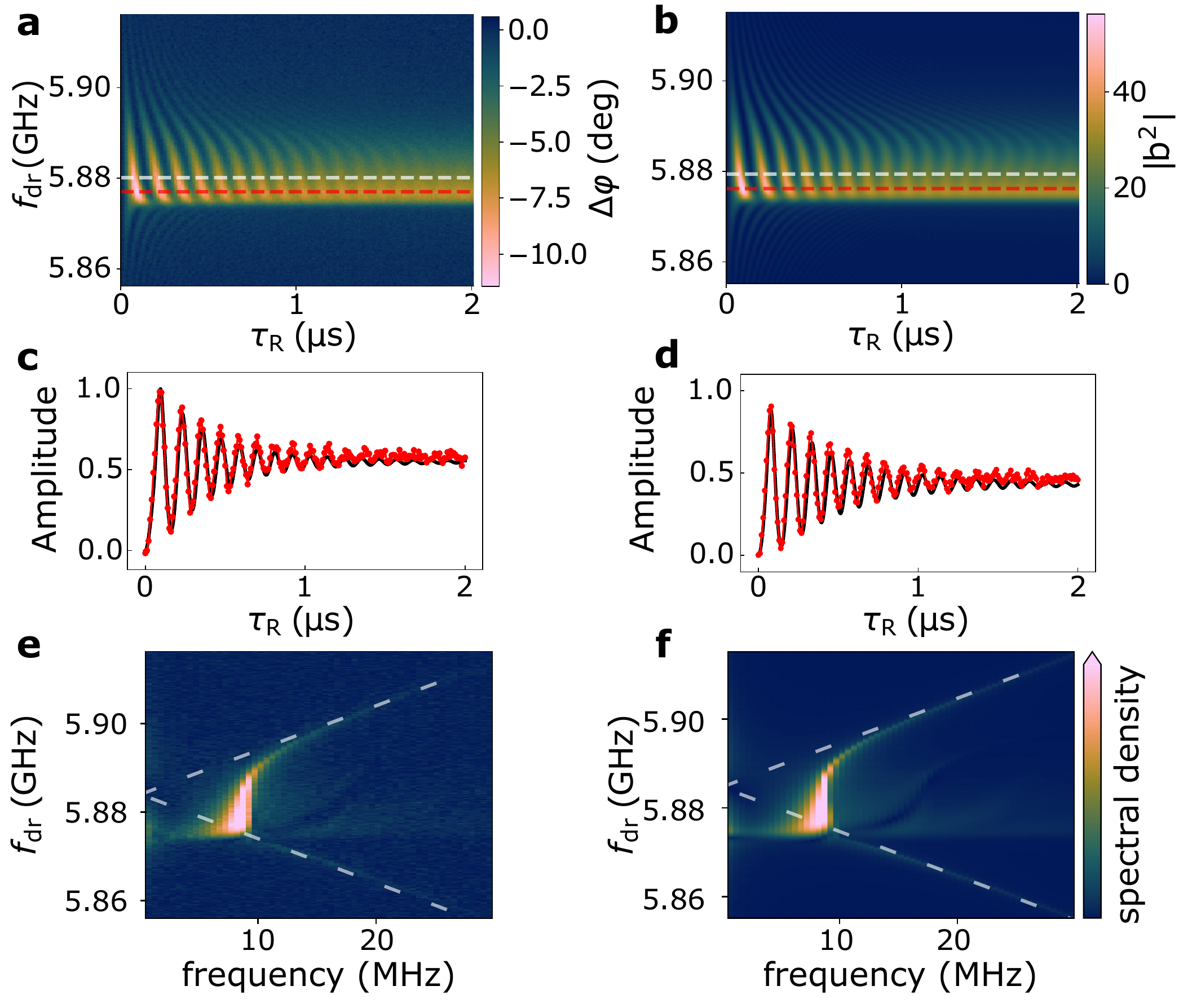}
\caption{\textbf{a.} Rabi chevrons of the grAl transmon at $B=\qty{0}{\tesla}$ as observed from the phase of the cavity signal at $f_R=\qty{6.0}{\giga\hertz}$. \textbf{b.} Modeling of the Rabi chevrons for $K=-2\pi \times \qty{200}{\kilo\hertz}$ and $\Gamma = 2\pi \times \qty{0.954}{\mega\hertz}$ and a Hilbert space truncated at $120$ quanta. The driving frequency is fixed using the frequency of the maximum dephasing of the measured chevron. \textbf{c.} Measurement (red dots) and modeling (black lines) for the Rabi-like damped oscillations of the grAl transmon for the red cuts represented in panels \textbf{a} and \textbf{b}. \textbf{d.} Measurement (red dots) and modeling (black lines) for the Rabi-like damped oscillations of the grAl transmon for the white cuts represented in panels \textbf{a} and \textbf{b}. \textbf{e.} Experimental Fourier transform of the Rabi chevrons showing the spectral content of the observed oscillations. \textbf{f.} Modeling of the Fourier transform of the Rabi chevrons showing good agreement with the experiment.
}
\label{Rabichevrons}%
\end{figure}

We first study Rabi oscillations as a function of the time $\tau_R$ of the excitation burst and the carrier frequency $f_{dr}$, at zero external magnetic field $B$ (see figure \ref{Rabichevrons}a). The state of the circuit is measured using the conventional dispersive readout i.e. the phase $\varphi$ of the microwave signal at the cavity resonance frequency $f_\mathrm{cav}=\qty{6.0}{\giga\hertz}$ here. The phase contrast $\Delta\varphi$ is simply related to the population of the circuit via $\Delta \varphi=2\chi \langle \hat{b}^{\dag} \hat{b}\rangle / \kappa$ where $\chi$ is the cross-Kerr coupling between the cavity photons and the circuit, $\kappa$ is the cavity linewidth and $\hat{b}$ is the transmon annihilation operator of the weakly anharmonic oscillator picture for our circuit. Defining $g$ as the circuit-cavity coupling strength, we have $\chi=-2 K g^2/(f_\mathrm{cav}-f_{0})^2$. The specific power of the drive and the cavity readout pulses are respectively $P_{dr}=-78\;$dBm and $P_{cav}=-103$ dBm. As explained above, the phase contrast is a direct measurement of $\langle \hat{b}^{\dag} \hat{b}\rangle$. As shown in figure \ref{Rabichevrons}a, we observe the characteristic fringes witnessing Rabi oscillations for conventional qubits. Nevertheless, we should stress that this is different regime from the qubit regime. This is first qualitatively shown by the observed unusual "Rabi chevron" pattern. Two features are particularly striking: first, the "chevron pattern" seems truncated below the continuous wave resonance frequency ; second, a closer inspection of the oscillations shows that they are non-sinuso\"{i}dal. Interestingly, the asymmetry can be qualitatively recovered by stating that, similarly to the two level system case, the chevrons result from a multiplication of the frequency envelope of the transition and the time domain damped Rabi oscillations. In such a heuristic picture, the asymmetry stems simply from the asymmetry of the lineshape of a Duffing oscillator.

It is possible to numerically model the dynamics of the circuit using as input parameters $K$, $\Gamma$ and the driving amplitude of the circuit thanks to the Qutip Python package. The simulation is run in the rotating frame by varying the detuning $\Delta f_\mathrm{dr}$ between the drive and $\omega_0$. The result of such a simulation is shown in figure \ref{Rabichevrons}b, where the driving frequency has been fixed using the frequency of the point with maximum contrast in the experiment. We find an excellent agreement with the data for $K=2\pi \times \qty{200}{\kilo\Hz}$ and $\Gamma = 2\pi \times \qty{0.954}{\mega\Hz}$. These values are close to those determined independently by the continuous wave spectroscopy of the circuit in figure \ref{grAl transmon}c. Interestingly, the population $\langle \hat{b}^{\dag} \hat{b}\rangle$ of the transmon which is encoded in the phase contrast is found to oscillate between nearly $0$ to about $60$. This shows that we are not in a qubit regime although we can observe quantum oscillations of the state of our circuit. The agreement between the modeling and the data is further exemplified in figure \ref{Rabichevrons}c and d. In order to further test the validity of the numerical simulation for modeling our circuit, it is interesting to Fourier transform the time domain signals. This is shown in figures \ref{Rabichevrons}e and f (for the experiment and the theory respectively). We observe again an excellent agreement and all the features observed experimentally are observed in the modeling. In particular, we identify a linear behavior at large positive or negative detuning. This is a sinuso\"{i}dal regime with a simple broadening arising from the decay rate of the circuit. In addition, the general features are strongly asymmetric in the detuning (vertical) axis. Such a behavior is not expected for a spin $1/2$ or a conventional transmon qubit as it should be symmetric and scale like $\Omega_{qubit}=\sqrt{\Omega_0^2+(\omega_{dr}-\omega_0)^2}$, where $\omega_{dr}= 2\pi f_{dr}$ and $\Omega_0$ is a linear function of the drive amplitude $\delta$. This is one important qualitative difference with our circuit. Close to \qty{5.88}{\giga\Hz}, we observe a broad feature in the spectrum indicating a non-sinuso\"{i}dal behavior. Rabi-like oscillations are observed here in a regime totally different than conventional qubits.

The observed "Rabi chevron" pattern is specific to the weakly anharmonic regime. It is interesting to put our findings in perspective with recent works using e.g. 2D material as a weak link for replacing the Josephson junction \cite{Kroll:18,Wang:19,Wang:22}. In reference \cite{Wang:19} in particular, there is a striking similarity of the observed "Rabi chevron" patterns with figure \ref{Rabichevrons}a. The immediate consequence of our findings is that it seems difficult to distinguish between the extreme weak anharmonic regime and the qubit regime simply by measuring individual time-domain cuts at a given detuning. We will see below that this holds true even if Ramsey-like fringes are observed.

\begin{figure}[!ht]
\centering\includegraphics[height=0.55\linewidth,angle=0]{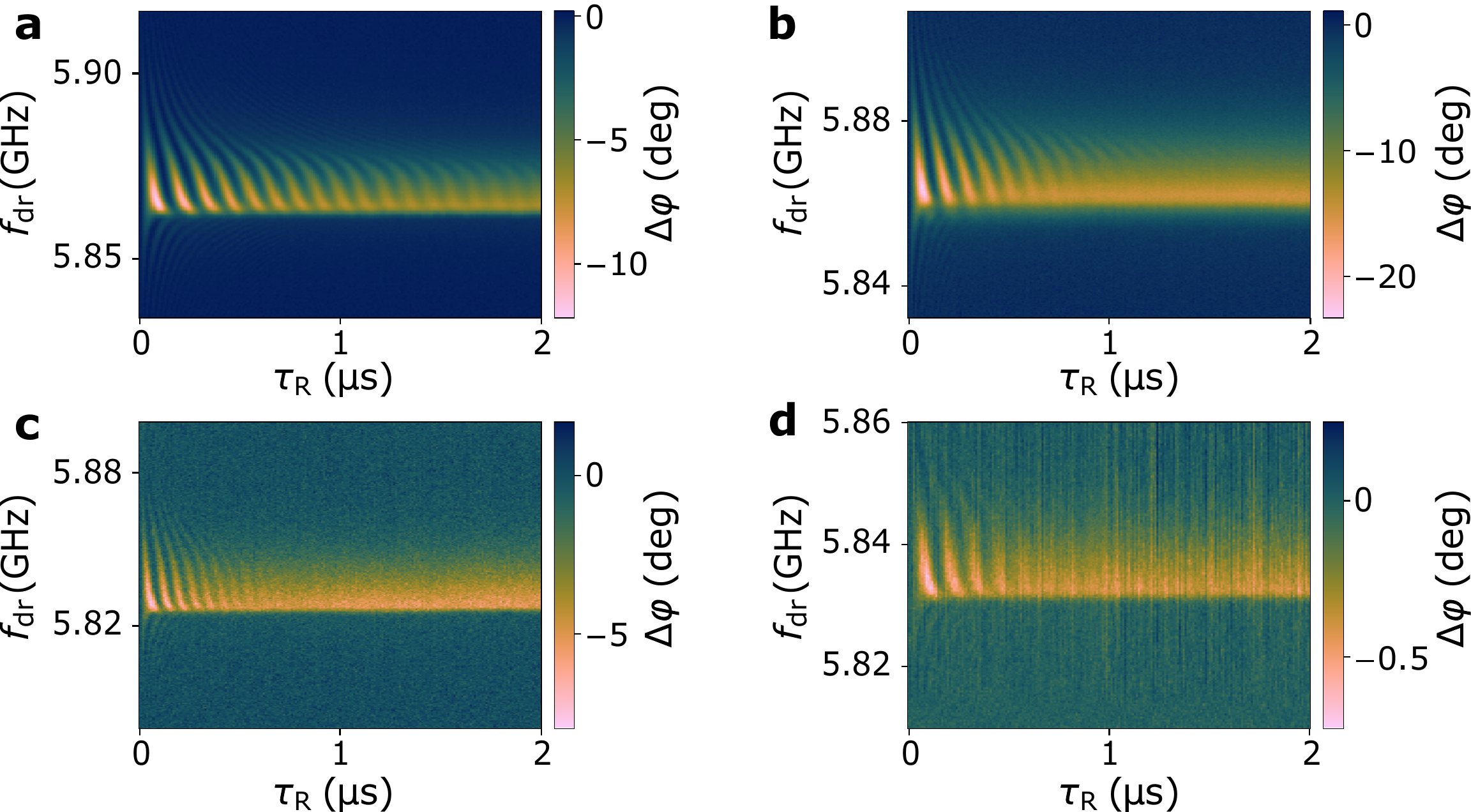}
\caption{\textbf{a.} Rabi chevrons of the grAl transmon at $B=\qty{0.15}{\tesla}$ \textbf{b.} Rabi chevrons of the grAl transmon at $B=\qty{0.45}{\tesla}$ \textbf{c.} Rabi chevrons of the grAl transmon at $B=\qty{0.6}{\tesla}$ \textbf{d.} Rabi chevrons of the grAl transmon at $B=\qty{0.7}{\tesla}$.
}%
\label{Bresilience}%
\end{figure}

The fact that the superconducting reservoirs are made of Nb, which has a rather high critical field, combined by the magnetic field resilience of GrAl films makes it possible to operate our superconducting circuits at high magnetic field. This is shown in figure \ref{Bresilience}. Each of the panels corresponds to a different magnetic field ranging from \qty{0.15}{\tesla} to \qty{0.7}{\tesla}. Whereas the Rabi chevrons are barely affected at \qty{0.15}{\tesla}, they are more affected at \qty{0.7}{\tesla}, but still clearly visible. The decoherence time decreases from $ T_{Rabi}\approx \qty{1.25}{\micro\s}$ to $T_{Rabi}\approx \qty{0.3}{\micro\s}$. We attribute this decrease to the slight misalignment between the B-field axis and the plane of the quantum circuit. Interestingly, we find a B-field resilience up to \qty{0.2}{\tesla} in out of plane B-field \cite{suppmat}. This is about two orders of magnitude larger than conventional superconducting qubits or grAl based resonators\cite{Borisov:20}.

How quantum are our observed oscillations if we are not in a qubit limit? This question echoes with the old debate on whether a weak anharmonic oscillator displaying Rabi oscillations can at all be considered as a quantum system or not \cite{Cirillo:05,Claudon:04,Claudon:08}. In our case, the Rabi chevron pattern is a strong discriminator between the classical and the quantum model which has an excellent agreement with our experiment. In particular, we can compare the expected frequency content between the two models. In the classical model, a single Rabi frequency is expected $\Omega_{cla}=\mathcal{A}(\delta)|\omega_{dr}-\omega_0|$, with $\mathcal{A}(\delta)$ a sublinear function of the driving amplitude $a$. Such an expression contrasts with the qubit case $\Omega_{qubit}$ which contains in addition a "transverse Rabi field" arising from the driving of the effective two level system in the rotating frame. The expression of $\Omega_{cla}$ means that we should expect only two symmetric lines in the maps \ref{Rabichevrons}e,f. While at large detuning around \qty{5.87}{\giga\Hz} or \qty{5.90}{\giga\hertz} a linear dispersion is present, this is clearly not the case at low detuning both in terms of frequency content and dispersion, as highlighted by the dashed lines in figure \ref{Rabichevrons}e,f. Thus, our findings are not explained by the classical anharmonic model. This further validates that we observe quantum oscillations in the time domain manipulation of our transmon circuit.

\begin{figure}[!ht]
\centering\includegraphics[height=0.75\linewidth,angle=0]{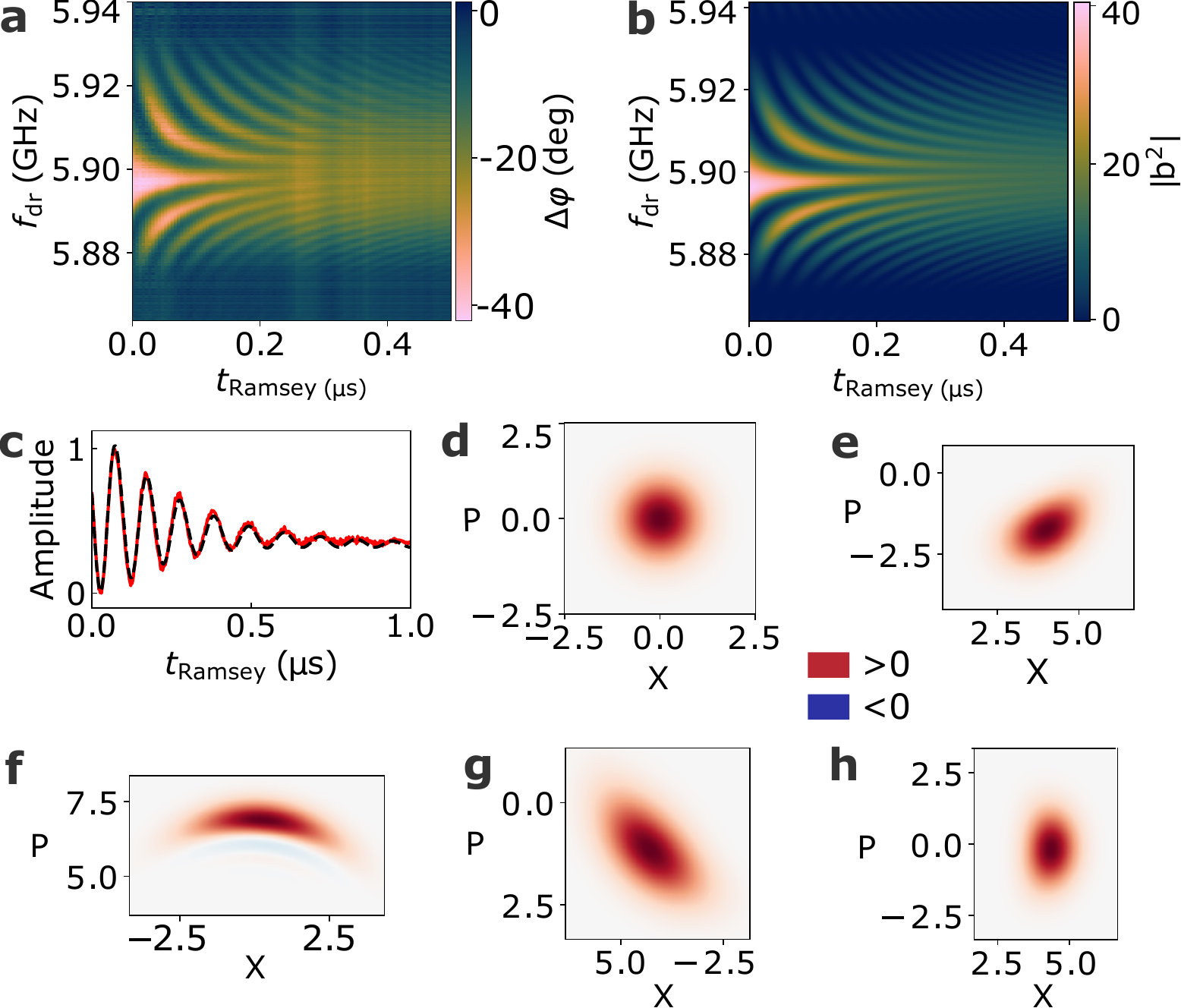}
\caption{\textbf{a.} Ramsey fringes as a function of detuning and time delay between the two $"\pi/2"$ pulse of duration $\tau_{0}$. \textbf{b.} Simulated Ramsey fringes for the same parameters as in figure \ref{Rabichevrons} and a Hilbert space truncated at 80 quanta. The duration of the $\pi/2$ pulses is \qty{31}{\nano\s}. The frequency is defined relatively to the measurement frequency. \textbf{c.} Measurement (red) and modeling (black)for the Ramsey interferometry of the grAl transmon after a $\pi/2$ pulse. The measurement pulse is \qty{20}{\nano\s} long at a frequency of \qty{5.91}{\giga\Hz}, the simulation pulse is \qty{21}{\nano\s} long at \qty{5.9016}{\giga\Hz}.  \textbf{d.,e.,f.,g.,h.} Simulated Wigner tomography for our parameters and $f_{dr}=\qty{5.9188}{\giga\Hz}$ for $t_{pulse}=0,\tau_{0},2\tau_{0},3\tau_{0}, 70\tau_{0}$.
}%
\label{Ramsey}
\end{figure}

The final picture for the dynamics of our system can be sketched by performing Ramsey interferometry. A first pulse of duration $\tau_{0}=\qty{30}{\nano\s}$ with $P_{dr}=-78\;\mathrm{dBm}$ is followed by a waiting time $\tau_\mathrm{Ramsey}$, by a second $\tau_{0}$ with $P_{dr}=-78\;\mathrm{dBm}$ and by a readout pulse. As in the conventional Ramsey interferometry setup, they correspond to half of one modulation in the Rabi sequence between the low and high number of quanta in figure \ref{Rabichevrons}c. As shown in the 2D map of figure \ref{Ramsey}a, we observe fringes which are the result of Ramsey interferences taking place in our circuit. Ramsey fringes are slightly asymmetrical, however, whereas there was a clear qualitative difference between the observed Rabi chevron pattern and that expected for a qubit, the asymmetry in the Ramsey fringes is not as noticeable. The difference in frequency between the Rabi chevrons of figure \ref{Rabichevrons} and the Ramsey fringes is due to a slight shift of the circuit frequency. The simulation of this experiment is shown in figure \ref{Ramsey}b. The duration of the pulses is \qty{31}{\nano\s}. The parameters are the same as in the Rabi chevrons simulation of figure \ref{Rabichevrons}, and the frequency offset is again set using the point of maximum contrast in the measurement. As for the Rabi chevrons, the modelling of the Ramsey interferometry reproduces closely the behaviour of the device. Ramsey oscillations after a pulse $\pi/2$ are shown in \ref{Rabichevrons}b for the measurement and the simulation. The frequency and length of the pulses are \qty{5.91}{\giga\Hz} and \qty{20}{\nano\s} for the measurement and \qty{5.9106}{\giga\Hz} and \qty{21}{\nano\s} in the simulation. We find an excellent agreement between the two curves. This agreement is also shown for time traces in figure \ref{Ramsey}c. As in the Rabi oscillations, they differ from the sinusoidal oscillations expected for Ramsey fringes in a 2 level system. We are led to conclude that the Rabi chevron pattern is more discriminant for disambiguating the qubit and the weak anharmonic regime although the quantitative analysis of both fringes reveals strong difference with the qubit case. Since we find a very good agreement between the modeling and our experimental findings, it is interesting to further simulate the Wigner tomography expected for different pulse times of our weakly anharmonic oscillator. Such a result is shown in figures \ref{Ramsey}c,d,e,f. We show the Wigner functions for pulse times $t_{pulse}=0,\tau_{0},2\tau_{0},3\tau_{0}, 70\tau_{0}$. Except from the starting point in vacuum, we see that there are clear signatures of non-classicality in the form of squeezing and negative parts in the Wigner tomography. Interestingly, the $2\tau_{0}$ leads to the most non-classical state as it displays negative forming a blue crescent in the Wigner function whereas $\tau_{0}$ displays simple squeezing. This is also a strong difference with the qubit case. In our system, the pulse time $\tau_{0}$ does not play a special role like for a qubit.

As a conclusion, we have investigated a simple superconducting circuit made out of a granular aluminium Josephson junction in the time domain. We observe quantum oscillations i.e. Rabi oscillations and Ramsey fringes. We have a quantitative understanding of the Rabi oscillations in the extreme weak anharmonic regime and the Ramsey fringes are qualitatively similar to the qubit regime. Although our quantum circuit is not a quantum bit, it displays all the requested features for quantum sensing and quantum amplification. In particular, we could in principle use it in the photon number resolved regime to use it as a single photon detector. Such detectors have been put forward recently for quantum sensing of cosmological objects using conventional transmon qubits. However, a major hurdle for the use of these single photon detector is magnetic field resilience. For example, there is a strong motivation to sense single microwave photon for quantum enhanced axion dark matter search \cite{Dixit:21} in cavities in strong magnetic field. Our quantum circuit could therefore have interesting applications for quantum sensing of cosmological signals.

\begin{acknowledgments}
We acknowledge fruitful discussions with I. Pop, A.A. Clerk, U. R\'eglade and Z. Leghtas. This work is supported by the ANR "MITIQ" and JCJC "STOIC", the Emergence project "MIGHTY" and the BPI project "QUARBONE".
\end{acknowledgments}


\begin{thebibliography}{99}
\bibitem{cavity}The Rabi chevrons at \qty{0.7}{\tesla} have been measured with a larger detuning $f_{cav}-f_{0}$, which explains the lower phase contrast.
\end{thebibliography}
\end{document}